\documentclass[twocolumn,showpacs,preprintnumbers,amsmath,amssymb,pre]{revtex4}
\usepackage{graphicx}% Include figure files
\usepackage{dcolumn}% Align table columns on decimal point
\usepackage{bm}% bold math
\usepackage{psfrag}

\begin{document}

\preprint{APS/123-QED}

\title{Convection in an ideal gas at high Rayleigh numbers}

\author{A. Tilgner}
%\email{andreas.tilgner@physik.uni-goettingen.de}
%\homepage{http://www.geo.physik.uni-goettingen.de/~atilgner}

\affiliation{Institute of Geophysics, University of G\"ottingen,
Friedrich-Hund-Platz 1, 37077 G\"ottingen, Germany }

\date{\today}% It is always \today, today,

\begin{abstract}
Numerical simulations of convection in a layer filled with ideal gas are
presented. The control parameters are chosen such that there is a significant
variation of density of the gas in going from the bottom to the top of the
layer. The relations between the Rayleigh, Peclet and Nusselt numbers depend
on the density stratification. It is proposed to use a data reduction which
accounts for the variable density by introducing into the scaling laws an
effective density. The relevant density is the
geometric mean of the maximum and minimum densities in the layer. A good fit to
the data is then obtained with power laws with the same exponent as for
fluids in the Boussinesq limit. Two relations
connect the top and bottom boundary layers: The kinetic energy densities
computed from free fall velocities are equal at the top and bottom, and the
products of free fall velocities and maximum horizontal velocities are equal for
both boundaries.
\end{abstract}

% PACS, the Physics and Astronomy Classification Scheme.
\maketitle

\section{Introduction}

There is by now a large body of literature on turbulent Rayleigh-B\'enard convection
\cite{Ahlers09}. Most of this
work strives to stay in a limit described by an approximation developed by
Oberbeck and Boussinesq, commonly known as the Boussinesq approximation \cite{Tritto88}.
Within this approximation, the material constants
are uniform across the layer and the temperature difference between top
and bottom boundaries is small compared with the absolute temperature at any
point in the layer. This idealized system has served as a paradigm for more
complicated convection problems, as for example atmospheric convection. A
major difference between convection in the Boussinesq limit and convection in
the atmosphere is that the latter occurs in a layer in which the gas density
varies significantly, which implies concomitant variations of viscosity and
thermal diffusivity.

Experiments on convection beyond the Boussinesq approximation have mostly focused on
effects caused by the temperature dependence of the material 
properties \cite{Wu91, Ahlers07}, as opposed to the effects of compressibility. 
Experiments in low temperature gases near their critical point are to some
extent an exception because the parameters
in these experiments can be adjusted such that the adiabatic temperature
gradient in the gas delays the onset of convection and shifts the critical
Rayleigh number by a detectable amount \cite{Ashken99,Kogan99,Burnis10}.
However, these experiments were restricted to convection near the onset.
Experimental studies aimed at turbulent Boussinesq convection in low temperature
gases also have to correct for the effect of the adiabatic temperature gradient
\cite{Chavan97,Niemal00}.

The present paper investigates through numerical simulation convecting ideal gas
in a layer with significant density variation from top to bottom. No
slip top and bottom boundaries are employed, so that the results are in principle
amenable to verification by laboratory experiments if one finds a way of
realizing similar density gradients in turbulent convection experimentally.
The parameters controlling the
density variation and the adiabatic temperature gradient are chosen to scatter
around the values of the terrestrial troposphere. The troposphere is the bottom
layer of the atmosphere of approximately 10 km thickness which is well mixed by
convection and bounded from above by the stably stratified stratosphere
\cite{Salby96}.

A long standing question in turbulent Rayleigh-B\'enard convection has been how
the heat transport across the layer depends on the control parameters. One goal
of the present paper will be to find out whether known results on convection in
an incompressible medium can be extended to an ideal gas. From a fundamental
point of view, the scale height of the density profile introduces a new length
scale into the problem in addition to the height of the layer. Previous studies
\cite{Toomre90} suggest that convective motion extends through multiple scale
heights, so that we cannot expect that the layer height will simply drop out of
the list of the relevant parameters in order to be replaced by the scale height.

Another issue peculiar to non-Boussinesq convection is the asymmetry between the
boundary layers next to the top and bottom boundaries. For example, the heat
conductivities near the warm bottom and cold top boundaries are different, but
the heat fluxes through both boundaries are identical in a statistically
stationary state, so that the temperature gradients within the top and bottom
boundary layers must be different. Because of this asymmetry, the temperature in
the center of a convection cell does not need to be equal to the arithmetic mean of top
and bottom boundary temperatures. The deviation of the true center temperature
from this arithmetic mean has been used in experiments as an indicator of
non-Boussinesq effects \cite{Wu91, Ahlers07}.

A typical situation in
astro- and geophysics is that only part of a convective layer is accessible to
observation, at least to accurate observation. For example, the top of planetary
or stellar atmospheres and the bottom of Earth's troposphere are better known
than the rest of the convective layers. It is in these cases important to know
what can be inferred about the convective layer from the observation of some
part of it. Translated to the idealized system simulated here, the question
arises as to what can be deduced about the whole convective layer or a boundary 
layer from knowledge of the opposite boundary layer.

The next section will present the mathematical model and the numerical method
used to solve it. The numerical results are analyzed in the third section,
in which one subsection deals
with the scaling of heat transfer and kinetic energy with the control
parameters, whereas another subsection is concerned with the relationship
between the boundary layers.

\section{Mathematical Model and Numerical Methods}

Consider a plane layer of height $d$ bounded by two planes perpendicular to the
$z-$axis. Gravity $\bm g$ is constant and pointing along the negative $z-$axis, $\bm g = -g
\hat{\bm z}$, where the hat denotes a unit vector. The ideal gas is
characterized by constant heat capacities at fixed volume and pressure, $C_V$
and $C_p$, and constant dynamic viscosity $\mu$ and heat conductivity $k$. This
implies that the density dependences of kinematic viscosity $\nu$ and thermal
diffusivity $\kappa$ are given by $\mu=\rho \nu$ and $k=\kappa \rho C_p$, in
which $\rho$ is the density. Let us assume that top and bottom boundaries are no
slip and have prescribed temperatures. Parameters evaluated at the top boundary
in the initial state
will be denoted by an index $o$ for ``outer''. The gas at the top of the layer
thus has temperature $T_o$. In the state specified by the initial conditions
(see Eq. (\ref{eq:density_init}) below), it also has kinematic viscosity $\nu_o$, thermal diffusivity
$\kappa_o$ and density $\rho_o$. The temperature difference across the layer is
$\Delta T$.

The system of equations governing density, temperature $T+T_o$, pressure $p$ and
velocity $\bm v$ reads with the usual summation convention over repeated
indices:

\begin{equation}
\partial_t \rho + \nabla \cdot (\rho \bm v) = 0
\label{eq:conti_dim}
\end{equation}

\begin{equation}
\rho [\partial_t \bm v +(\bm v \cdot \nabla) \bm v]= -\nabla p +\rho \bm g +
\mu [\nabla^2 \bm v + \frac{1}{3} \nabla (\nabla \cdot \bm v)]
\label{eq:NS_dim}
\end{equation}

\begin{equation}
\partial_t T + \bm v \cdot \nabla T =\frac{C_p}{C_V}\kappa \nabla^2 T -
\frac{p}{\rho C_V} \nabla \cdot \bm v + \frac{2\mu}{\rho C_V} [e_{ij}-\frac{1}{3} (\nabla \cdot
\bm v) \delta_{ij}]^2
\label{eq:T_dim}
\end{equation}

\begin{equation}
p=\rho R (T+T_o)
\label{eq:state}
\end{equation}

The gas constant $R$ in the equation of state is given by $ R=R_u/m $, with the
molar mass $m$ and the universal gas constant $R_u=8.314 J~ mol^{-1}~ K^{-1}$.
It follows from thermodynamics that $R=C_p-C_V$. The strain rate tensor $e_{ij}$
is given by $e_{ij}=\frac{1}{2}(\partial_j v_i + \partial_i v_j)$.

From here on, we will use nondimensional variables. All lengths are expressed in
multiples of $d$, and the scales of time and density are chosen as $d^2/\kappa_o$
and $\rho_o$, respectively. The difference between the temperature of the gas
and the top temperature, $T$, is scaled with $\Delta T$. Using the same symbols
for the non-dimensional variables space, time, density, velocity and temperature
difference with the top boundary as in the dimensional equations
(\ref{eq:conti_dim}-\ref{eq:state}), one obtains the system

\begin{equation}
\partial_t \rho + \nabla \cdot (\rho \bm v) = 0
\label{eq:conti}
\end{equation}

\begin{eqnarray}
\partial_t \bm v +(\bm v \cdot \nabla) \bm v & = &
-\frac{1}{\rho}\nabla [(T+\frac{T_o}{\Delta T})\rho]\frac{1}{\gamma}
\frac{H_o}{d} \mathrm{Pr} ~ \mathrm{Ra} 
\label{eq:NS}
\\
& - & \hat{\bm z}\mathrm{Pr} ~ \mathrm{Ra}\frac{T_o}{\Delta T}+
[\nabla^2 \bm v + \frac{1}{3} \nabla (\nabla \cdot \bm v)]\frac{1}{\rho}\mathrm{Pr}
\nonumber
\end{eqnarray}

\begin{eqnarray}
\partial_t T + \bm v \cdot \nabla T & = & \frac{\gamma}{\rho}\nabla^2 T -
(\gamma-1)(T+\frac{T_o}{\Delta T}) \nabla \cdot \bm v 
\label{eq:T}
\\ 
& + & [e_{ij}-\frac{1}{3} (\nabla \cdot \bm v) \delta_{ij}]^2 \frac{1}{\rho} 2 \gamma
(\gamma-1)
\frac{d}{H_o} \frac{1}{\mathrm{Ra}}
\nonumber
\end{eqnarray}

together with the boundary conditions

\begin{equation}
T(z=1)=0 ~~~,~~~ T(z=0)=1 ~~~,~~~ \bm v(z=1)= \bm v(z=0) =0.
\label{eq:bc}
\end{equation}
The equation of state has been used to eliminate pressure.

Seven parameters control the system. The Rayleigh number $\mathrm{Ra}$ is the
usual Rayleigh number evaluated at the top boundary (remember that the thermal
expansion coefficient of an ideal gas is its inverse temperature):
\begin{equation}
\mathrm{Ra} = \frac{g d^3 \Delta T}{T_o \kappa_o \nu_o}.
\end{equation}
The Prandtl number $\mathrm{Pr}$ is independent of space in the present model
and is set to 0.7 in all calculations:
\begin{equation}
\mathrm{Pr} = \frac{\nu}{\kappa} =0.7.
\end{equation}
The adiabatic exponent $\gamma$ is set in this paper to its value for a
monoatomic gas:
\begin{equation}
\gamma=C_p/C_V=\frac{5}{3}.
\end{equation}
The density stratification is specified by $d/H_o$, where $H_o$ is the
adiabatic scale height at the top boundary, $H_o=\gamma R T_o/g$. The meaning of the
fifth parameter, $\Delta T/T_o$, is obvious from the definitions above. An
alternative parameter, redundant after the choices made so far, is the ratio of
the adiabatic temperature difference between top and bottom, $\Delta
T_{\mathrm{ad}}$, and the actual temperature difference, $\Delta T$:
\begin{equation}
\frac{\Delta T_{\mathrm{ad}}}{\Delta T} = (\gamma-1)\frac{T_o}{\Delta
T}\frac{d}{H_o}
= \frac{gd}{C_p \Delta T}
\label{eq:Delta_T_ad}
\end{equation}
This ratio needs to be less than 1 for any convection to occur.

The sixth ``parameter'' is the initial temperature and density distribution. The
initial conditions appear as a control parameter because they specify for
instance the total mass in the layer. They also determine 
$\rho_o$ and hence $\nu_o$ and $\kappa_o$,
a quantity used to make the governing equations nondimensional. All
simulations are started from zero velocity and the conductive profile $T=1-z$.
The density is then determined from (\ref{eq:NS}):
\begin{equation}
\rho = \rho_o \left( \frac{1}{1+\frac{1-z}{T_o/\Delta T}} \right) ^{1-\gamma
\frac{d}{H_o}\frac{T_o}{\Delta T}}
\label{eq:density_init}
\end{equation}

The geometry, quantified through the aspect ratio of the computational volume,
is the seventh parameter. Periodic boundary conditions are imposed in $x-$ and
$y-$directions with periodicity lengths $l_x$ and $l_y$. All computations have
been made for $l_x=l_y=2d$. No aspect ratio dependence has been investigated
since the main interest of the present work was to determine the effects of
density variations.

Even though it was chosen to keep several parameters fixed, there still remains
a vast parameter space to explore since $\mathrm{Ra}$, $d/H_o$ and $\Delta
T/T_o$ have to be varied. The computations below are roughly guided by the
terrestrial troposphere \cite{Salby96}, for which $\Delta T/T_o \approx 0.35$ and
$d/H_o \approx 1.9$. The density of air varies by a factor between 3 and 4 within
the troposphere. Note that $\gamma=1.4$ is the appropriate adiabatic exponent for
air.

\begin{figure}
\includegraphics[width=8cm]{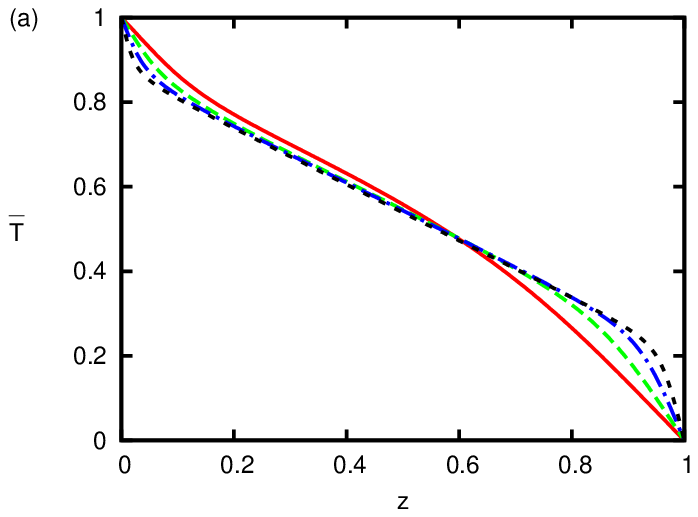}
\includegraphics[width=8cm]{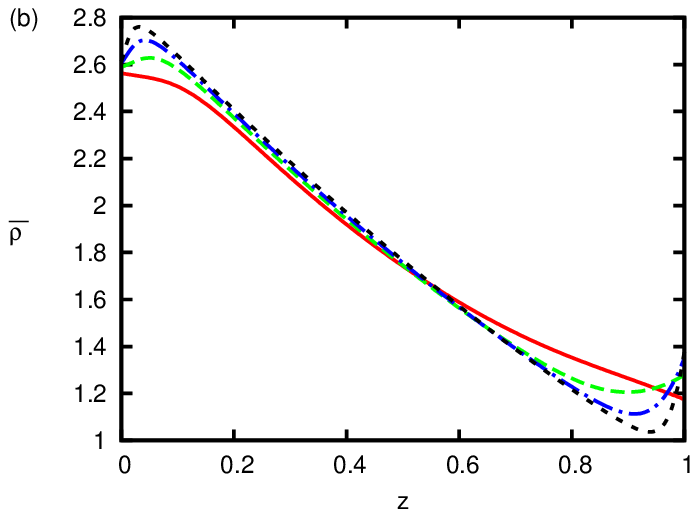}
\includegraphics[width=8cm]{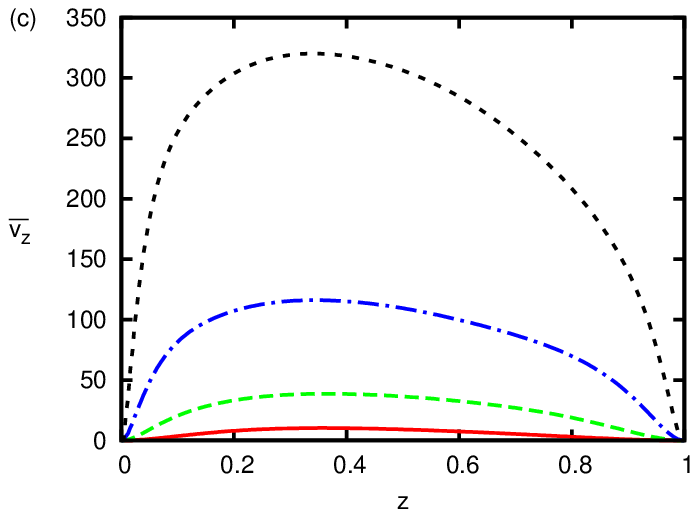}
\includegraphics[width=8cm]{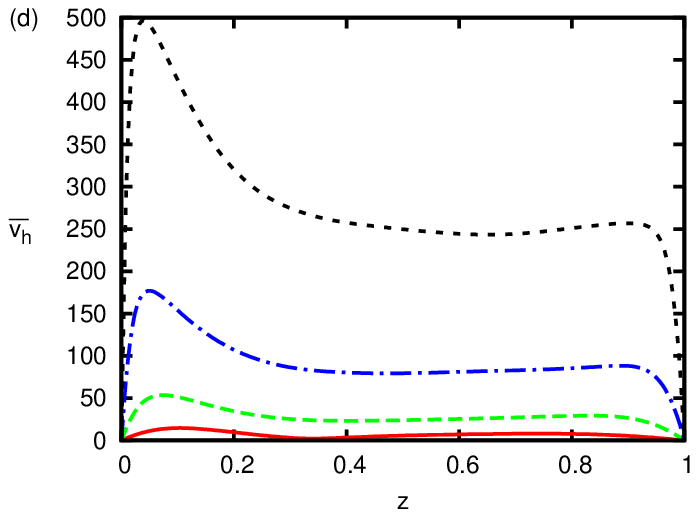}
\caption{(Color online)
Variations as a function of height $z$ of horizontal averages of temperature
$T$ (top panel), density $\rho$ (second panel from the top), vertical velocity
$v_z$ (third panel) and horizontal velocity (bottom panel). The overbars signal
averages over time and horizontal planes. The different traces
are for $\frac{\Delta T}{T_o}=\frac{d}{H_o}=3$ and the Rayleigh numbers 
$2 \times 10^4$ (solid red line), $2 \times 10^5$ (long dashed green line), $2
\times 10^6$ (dot dashed blue line), $2 \times 10^7$ (short dashed black line).}
\label{fig:profile}
\end{figure}

The well known Boussinesq approximation is recovered from equations
(\ref{eq:conti}-\ref{eq:T}) in the limit $d/H_o \rightarrow 0$ and
$\Delta T/T_o \rightarrow 0$ \cite{Tritto88}. In this limit, the sound speed goes
to infinity. A purely explicit time step will be used in the numerical method
below, so that simulations close to the Boussinesq limit become impractical. For
this reason, and since the Boussinesq limit is an important reference case, a
second system of equations was also implemented numerically:
\begin{equation}
\partial_t \rho + \nabla \cdot \bm v = 0
\label{eq:conti_BQ}
\end{equation}

\begin{equation}
\partial_t \bm v +(\bm v \cdot \nabla) \bm v = - c^2 \nabla \rho +
\mathrm{Pr} ~ \mathrm{Ra} ~\theta {\bm z} + \mathrm{Pr} \nabla^2 \bm v
\label{eq:NS_BQ}
\end{equation}

\begin{equation}
\partial_t \theta + \bm v \cdot \nabla \theta -v_z =\nabla^2 \theta
\label{eq:T_BQ}
\end{equation}
where $\theta$ is the deviation from the conductive profile, $T=1-z+\theta$. The sound
speed $c$ appears as an independent variable. For small Mach numbers, the
density fluctuations are small and the equation of continuity can be linearized
to yield (\ref{eq:conti_BQ}). The Boussinesq equations are recovered in the
limit $c \rightarrow \infty$. In the simulations mentioned below using 
(\ref{eq:conti_BQ}-\ref{eq:T_BQ}), the sound speed was adjusted such that the
Mach number always stayed below 0.1. Note that simulations of weakly
compressible convection as an approximation to Boussinesq convection have been
undertaken before. For instance, simulations using the lattice Boltzmann method
\cite{Lohse03} implicitly do so.

Systems (\ref{eq:conti}-\ref{eq:T}) and (\ref{eq:conti_BQ}-\ref{eq:T_BQ}) have
been simulated with a finite difference method implemented on graphic processing
units using C for CUDA. The numerical method used centered finite differences of
second order on a collocated grid except for the advection terms which used an
upwind biased third order scheme. The time step was a third order Runge-Kutta
method. The standard resolution was $256^3$. Lower resolution was sufficient at
the smallest simulated $\mathrm{Ra}$. The validation of the code is described in
the appendix.

\section{Results}

\subsection{Overview}
A summary of the simulations is given in table \ref{table1}. Apart from the
control parameters, it lists the Nusselt number, defined as
\begin{equation}
\mathrm{Nu}=-\overline{\frac{d T}{dz}}
\label{eq:Nu}
\end{equation}
where the overbar denotes average over time and either top or bottom boundary.
The kinetic energy density $E_{\mathrm{kin}}$ is given by
\begin{equation}
E_{\mathrm{kin}}=\frac{1}{V} \int \frac{1}{2} \rho \bm v^2 dV
\label{eq:Ekin}
\end{equation}
whereas the Peclet number $\mathrm{Pe}$ is computed from
\begin{equation}
\mathrm{Pe}=\sqrt{\frac{1}{V} \int \frac{1}{2} \bm v^2 dV}
\label{eq:Pe}
\end{equation}
which is aptly called Peclet number because velocities are computed in units of
$\kappa_o/d$. The average temperature deviation from the conductive profile at
the center of the layer, $\bar \theta_m$, is also listed in table \ref{table1},
together with the average density in the midplane, $\bar \rho_m$.

Fig. \ref{fig:profile} shows vertical profiles of temperature, density, vertical velocity,
and horizontal velocity $(v_x^2+v_y^2)^{1/2}$ for different $\mathrm{Ra}$.
Contrary to the Boussinesq case, these profiles are not symmetric about the
midplane. The relation between the top and bottom regions of the layer will be
discussed in section \ref{BL}. As
$\mathrm{Ra}$ increases, an increasingly large interval develops in which the
temperature gradient is approximately equal to the adiabatic gradient. The
maximum of the vertical velocity is found below the midplane. Two local maxima show up 
in the profiles of horizontal velocity. The larger velocities are found near the
bottom boundary. When both $d/H_o$ and $Ra$ are large, there is no local maximum
of horizontal velocity near the top boundary and the horizontal velocity
decreases monotonically with height. These cases result in blanks in the
last column of table \ref{table1}. Boundary layers
also exist in the density profiles. The average density takes its maximum and
minimum values, $\rho_\mathrm{max}$ and $\rho_\mathrm{min}$, near the bottom and
top boundary, respectively. Both $\rho_\mathrm{max}$ and $\rho_\mathrm{min}$ are 
given in table \ref{table1}, too. The ratio $\rho_\mathrm{max}/\rho_\mathrm{min}$ is
generally less than is suggested by the values of $d/H_o$ and the initial
conditions (\ref{eq:density_init}) because the statistically stationary,
turbulent and well mixed state is
nearly adiabatic, not conductive. A rough estimate of
$\rho_\mathrm{max}/\rho_\mathrm{min}$ can be obtained from assuming that the
adiabatic state extends throughout the layer, implying that the density takes
its extremal values exactly on the boundaries. This leads to
\begin{equation}
\frac{\rho_\mathrm{max}}{\rho_\mathrm{min}} \approx
\left( \frac{T_m+\frac{1}{2}\Delta T_\mathrm{ad}}{T_m-\frac{1}{2}\Delta
T_\mathrm{ad}} \right)
^{1/(\gamma-1)}
\end{equation}
with $T_m=(\bar \theta_m + \frac{1}{2}) \Delta T + T_o$,
which is compatible with the numerical results.

\begin{figure}
\includegraphics[width=8cm]{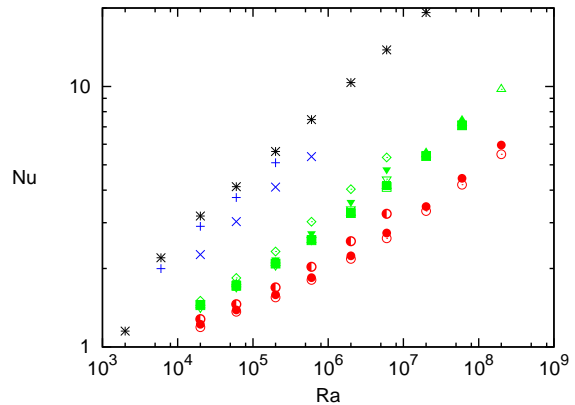}
\caption{(Color online)
Nusselt number $\mathrm{Nu}$ as a function of Rayleigh number $\mathrm{Ra}$
within the Boussinesq approximation (black stars), for 
$\Delta T_{\mathrm{ad}}/\Delta T = 1/15$ (blue symbols) and $\Delta T/T_o=0.1$
(plus) and 1 (x), for $\Delta T_{\mathrm{ad}}/\Delta T = 2/3$ (green symbols) and
$\Delta T/T_o=0.1$ (empty squares), 0.3 (full squares), 1 (empty triangle
up), 3 (full triangle up), 10 (empty triangle down), 30 (full triangle down) and
100 (empty diamonds)
and for $\Delta T_{\mathrm{ad}}/\Delta T = 4/5$ (red symbols) and
$\Delta T/T_o=0.1$ (empty circles), 1 (full circles) and 10 (half filled
circles). Data for the Boussinesq case do not appear in the subsequent figures.
}
\label{fig:Nu_Ra}
\end{figure}

\subsection{Global Quantities}
The most immediate task is of course to find predictions for the Nusselt number.
A straightforward plot of $\mathrm{Nu}$ vs. $\mathrm{Ra}$ (fig. \ref{fig:Nu_Ra})
shows that one does not obtain simple power laws for large $\Delta
T_{\mathrm{ad}}/\Delta T$ and $d/H_o$. A finite adiabatic temperature gradient
modifies the onset of convection. If one aims for a data reduction which
collapses the different curves in fig. \ref{fig:Nu_Ra}, one can account for the
adiabatic temperature difference by defining a corrected Rayleigh number
$\mathrm{Ra}_*$ by
\begin{equation}
\mathrm{Ra}_*=\frac{g d^3 (\Delta T- \Delta T_\mathrm{ad})}{T_o \kappa_o \nu_o}
=\mathrm{Ra} \left( 1-(\gamma-1) \frac{d}{H_o} \frac{T_o}{\Delta T} \right).
\label{eq:Ra_NB}
\end{equation}
A similar correction seems in order for $\mathrm{Nu}$. Since the adiabatic
temperature gradient needs to be established before any convection can start, it
is natural to subtract the heat conducted down the adiabat from both the actual
heat transport and the conductive heat transport used for the normalization of
the heat transport \cite{Massag80}:
\begin{equation}
\mathrm{Nu}_* = \frac{\mathrm{Nu} - \Delta T_\mathrm{ad}/\Delta T}{1-\Delta
T_\mathrm{ad}/\Delta T}.
\label{eq:Nu_NB}
\end{equation}

\begin{figure}
\includegraphics[width=8cm]{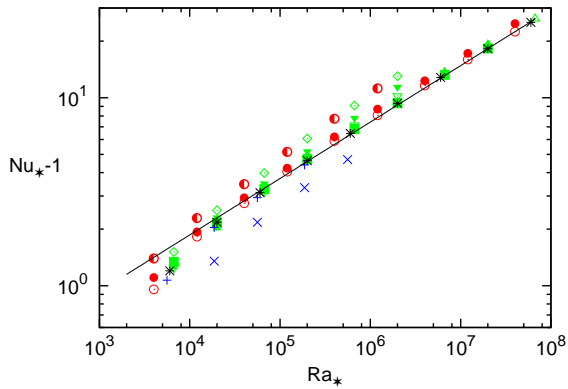}
\caption{(Color online)
$\mathrm{Nu}_*$ as a function of $\mathrm{Ra}_*$ with the same symbols as in
fig. \ref{fig:Nu_Ra}. The solid line indicates the power law
$\mathrm{Nu}_*=\mathrm{Ra}_*^{0.3}/8.5$.}
\label{fig:Nu_NB_Ra_NB}
\end{figure}

It is seen from fig. \ref{fig:Nu_NB_Ra_NB} that for $\mathrm{Ra}_*$ larger
than roughly $10^5$, one
finds approximately $(\mathrm{Nu}_*-1) \propto \mathrm{Ra}_*^{0.3}$, but the
prefactors depend on the other control parameters in a non-trivial way. This is
compatible with an argument exposed in Ref. \onlinecite{Furuka02}, which states that
$\mathrm{Ra} (\mathrm{Nu}-1) = f(\mathrm{Ra}_*)$ with $f$ an a priori unknown
function. The argument has to assume negligible viscous heating and small
density variations, so that it is not expected to hold throughout the parameter
range investigated here. Nonetheless, a best fit to the data yields
$\mathrm{Ra} (\mathrm{Nu}-1) =\mathrm{Ra}_*^{1.3}/8.5$
which can be rearranged into the fitting function in fig. \ref{fig:Nu_NB_Ra_NB}, 
$(\mathrm{Nu}_*-1) = \mathrm{Ra}_*^{0.3}/8.5$.

\begin{figure}
\includegraphics[width=8cm]{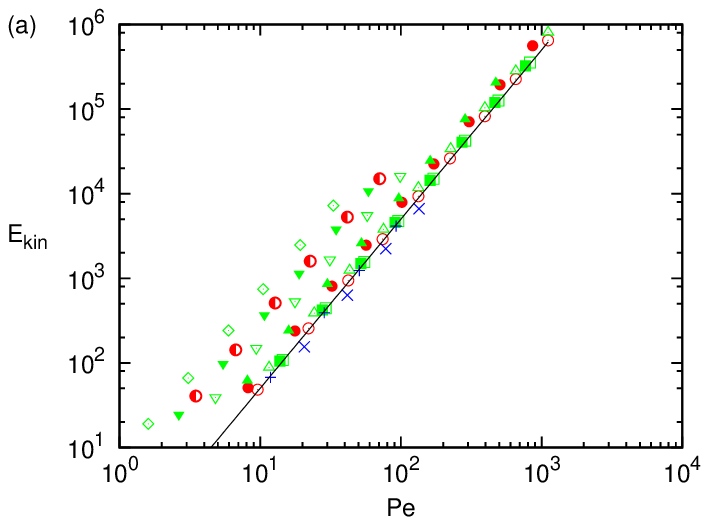}
\includegraphics[width=8cm]{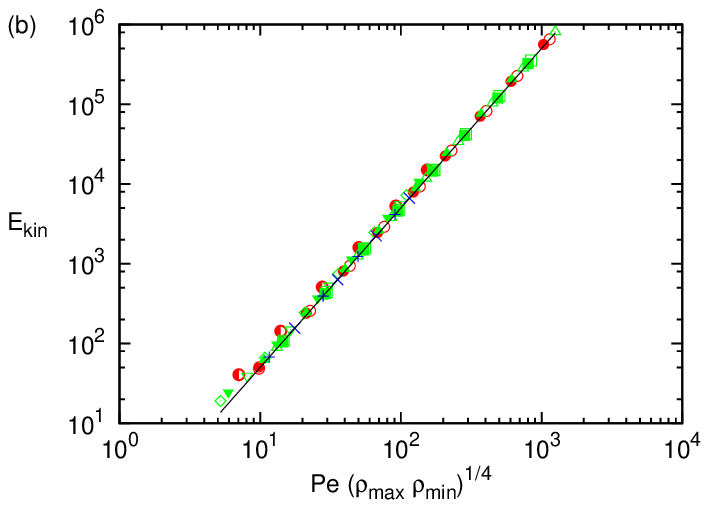}
\caption{(Color online)
The kinetic energy density $E_{\mathrm{kin}}$ as a function of $\mathrm{Pe}$
(top panel) and $\mathrm{Pe}(\rho_\mathrm{max} \rho_\mathrm{min})^{1/4}$ (bottom
panel) with the same symbols as in
fig. \ref{fig:Nu_Ra}. The solid lines are given by
$E_{\mathrm{kin}}=\frac{1}{2} \mathrm{Pe}^2$ (top panel) and
$E_{\mathrm{kin}}=\frac{1}{2} \sqrt{\rho_\mathrm{max} \rho_\mathrm{min}} ~
\mathrm{Pe}^2$ (bottom panel).}
\label{fig:Ekin_Pe}
\end{figure}

It can also be useful to relate $\mathrm{Nu}$ to the kinetic energy or the
Peclet number. It was noted in Ref. \onlinecite{Schmit09} that 
$(\mathrm{Nu}-1) \propto \mathrm{Pe}^{2/3}$ in Boussinesq convection in
computational volumes of large aspect ratio, which is equivalent to
$(\mathrm{Nu}-1) \propto E_{\mathrm{kin}}^{1/3}$ in that case. In the present
simulations, the relation between $E_{\mathrm{kin}}$ and $\mathrm{Pe}$ is
already non-trivial (see fig. \ref{fig:Ekin_Pe}),
because there is a factor representing an effective density
between the two quantities. It turns out that the geometric mean of
$\rho_\mathrm{max}$ and $\rho_\mathrm{min}$ is a suitable effective density to
the extent that in fig. \ref{fig:Ekin_Pe}, all points for
$\mathrm{Ra}_*(\rho_\mathrm{max}
\rho_\mathrm{min})^{1/4} > 100$ deviate by less
than 30 \% in $E_{\mathrm{kin}}$ from
\begin{equation}
E_{\mathrm{kin}}=\frac{1}{2} \sqrt{\rho_\mathrm{max} \rho_\mathrm{min}} ~
\mathrm{Pe}^2.
\end{equation}

\begin{figure}
\includegraphics[width=8cm]{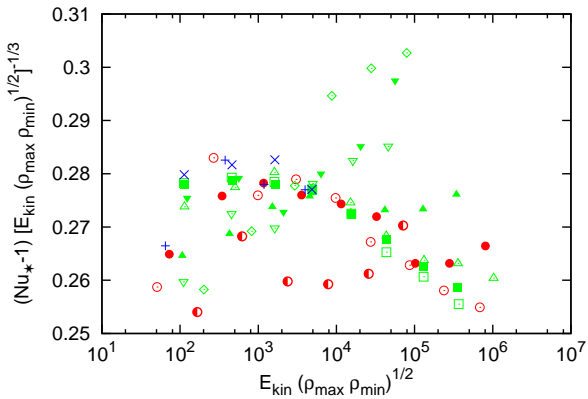}
\caption{(Color online)
$\mathrm{Nu}_*-1$ as a function of $E_{\mathrm{kin}}(\rho_\mathrm{max}
\rho_\mathrm{min})^{1/2}$, compensated for the power law in Eq.
(\ref{equ:Nu_NB_Ekin}), with the same symbols as in
fig. \ref{fig:Nu_Ra}.}
\label{fig:Nu_NB_Ekin}
\end{figure}

\begin{figure}
\includegraphics[width=8cm]{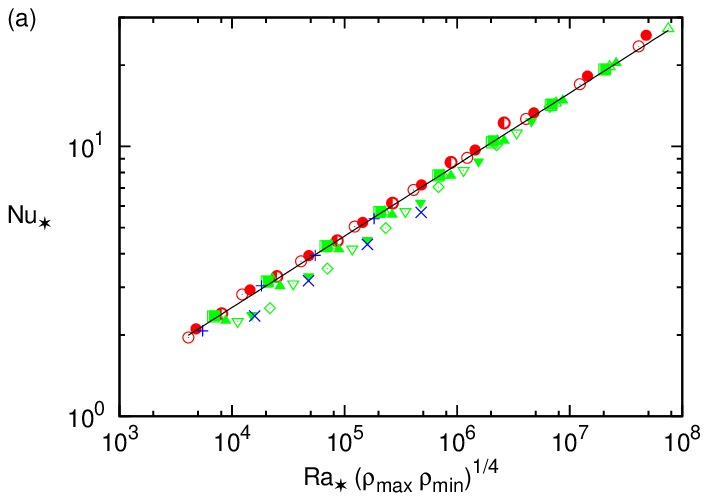}
\includegraphics[width=8cm]{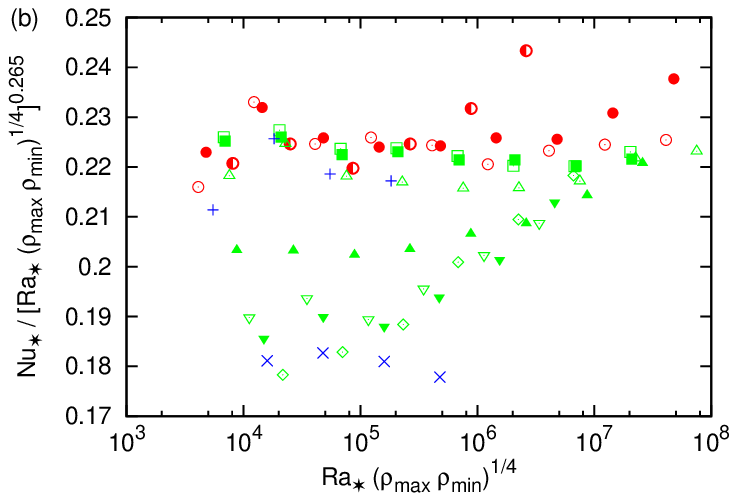}
\caption{(Color online)
$\mathrm{Nu}_*$ as a function of $\mathrm{Ra}_*(\rho_\mathrm{max}
\rho_\mathrm{min})^{1/4}$, on a double logaritmic scale in the top panel
and compensated for the power law in Eq.
(\ref{equ:Nu_NB_Ra_NB2}) in the bottom panel, with the same symbols as in
fig. \ref{fig:Nu_Ra}.}
\label{fig:Nu_NB_Ra_NB2}
\end{figure}

This geometric mean becomes again important when looking for a relation between
$\mathrm{Nu}_*-1$ and $E_{\mathrm{kin}}$. A good fit to the data is obtained
from
\begin{equation}
\mathrm{Nu}_*-1 = \frac{2}{7} (E_{\mathrm{kin}} \sqrt{\rho_\mathrm{max}
\rho_\mathrm{min}})^{1/3}
\label{equ:Nu_NB_Ekin}
\end{equation}
(see fig. \ref{fig:Nu_NB_Ekin}) which reduces of course to the previously known
scaling \cite{Schmit09,Schmit10} for $\rho_\mathrm{max}=\rho_\mathrm{min}$
and $\Delta T_\mathrm{ad}/\Delta T=0$.

Having established the relevance of the product $\rho_\mathrm{max}
\rho_\mathrm{min}$, it is tempting to introduce it into fits of $\mathrm{Nu}_*$
vs. $\mathrm{Ra}_*$. A reasonable fit is shown in fig. \ref{fig:Nu_NB_Ra_NB2} to be
\begin{equation}
\mathrm{Nu}_*= 0.22 \left( \mathrm{Ra}_* (\rho_\mathrm{max}
\rho_\mathrm{min})^{1/4} \right)^{0.265},
\label{equ:Nu_NB_Ra_NB2}
\end{equation}
which is an improvement compared with fig. \ref{fig:Nu_NB_Ra_NB} especially for
$\mathrm{Ra}_*(\rho_\mathrm{max}
\rho_\mathrm{min})^{1/4} > 10^6$.

\subsection{Boundary Layers \label{BL}}
Previous studies have quantified the asymmetry between top and bottom in
non-Boussinesq convection with the help of the midplane temperature \cite{Wu91,Ahlers07}. In most of
the present simulations, the midplane temperature deviates by less than 
$0.05~ \Delta T$ from its value in the conductive state (see table \ref{table1}).
Such a small deviation is difficult to determine accurately and requires long
time integrations, so that this section will not consider $\bar \theta_m$ any
further, apart from noting that $\bar \theta_m$ is negative for small 
$\Delta T_{\mathrm{ad}}/\Delta T$ (in agreement with Ref. \onlinecite{Ahlers07}) but
becomes positive for $\Delta T_{\mathrm{ad}}/\Delta T$ large enough.

A relation between temperature boundary layers deduced from experimental data by
Wu and Libchaber \cite{Wu91} is based on a temperature scale computed from
quantities local to each boundary layer. In many of the simulations presented
here, the boundary layers are still quite thick and there is significant
variation of for example thermal diffusivity across them, so that the results of
Wu and Libchaber cannot be tested in a meaningful way.

\begin{figure}
\includegraphics[width=8cm]{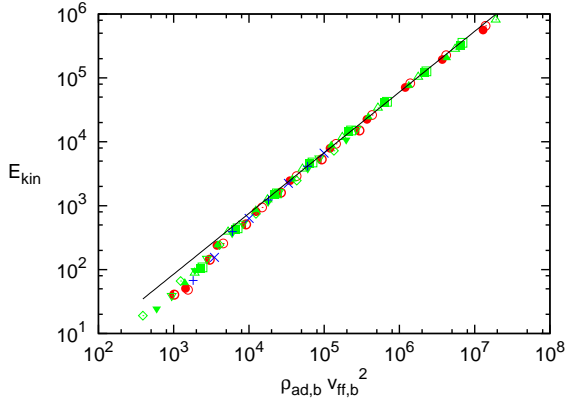}
\caption{(Color online)
$E_{\mathrm{kin}}$ as a function of $\rho_\mathrm{ad,b} v_\mathrm{ff,b}^2$
with the same symbols as in fig. \ref{fig:Nu_Ra}. The solid line indicates the power law 
$E_{\mathrm{kin}}=0.12 ~ (\rho_\mathrm{ad,b} v_\mathrm{ff,b}^2)^{0.95}$.}
\label{fig:Ekin_vff}
\end{figure}

\begin{figure}
\includegraphics[width=8cm]{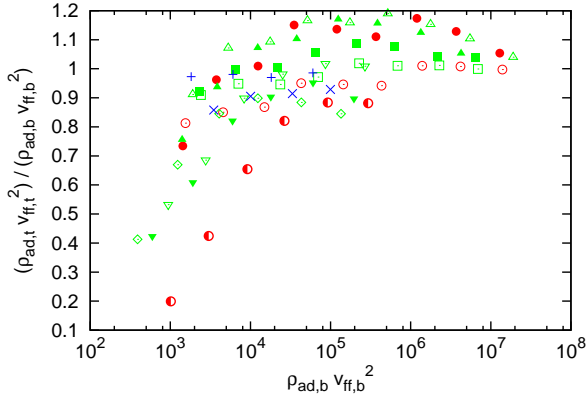}
\caption{(Color online)
$(\rho_\mathrm{ad,t} v_\mathrm{ff,t}^2)/(\rho_\mathrm{ad,b}
v_\mathrm{ff,b}^2)$ as a function of $\rho_\mathrm{ad,b}
v_\mathrm{ff,b}^2$ with the same symbols as in fig. \ref{fig:Nu_Ra}.}
\label{fig:vff_t_b}
\end{figure}

In the following, overbars denote averages over time and horizontal planes, and
the indices b and t indicate top and bottom boundaries. For example, $\bar
\rho_t$ is the average density at the top boundary. It is in general different
from $\rho_o$, which is the density at the top boundary in the initial
conductive state given by Eq. (\ref{eq:density_init}).

This subsection will present two relations between the top and bottom regions of
the convection layer involving the free fall velocity. In order to compute a
free fall velocity, we define $T_\mathrm{ad,t}$ and $T_\mathrm{ad,b}$
as the (dimensional) temperatures 
the gas would have at the top and bottom boundaries if the adiabatic temperature
profile extended throughout the entire layer:
\begin{equation}
T_\mathrm{ad,t}=T_m-\frac{1}{2} \Delta T_\mathrm{ad} ~~~,~~~
T_\mathrm{ad,b}=T_m+\frac{1}{2} \Delta T_\mathrm{ad}
\label{eq:T_ad}
\end{equation}
with $T_m=(\bar \theta_m + \frac{1}{2}) \Delta T + T_o$. Under the same
assumption, the densities at the two boundaries, $\rho_\mathrm{ad,t}$ and
$\rho_\mathrm{ad,b}$, are given by
\begin{equation}
\rho_\mathrm{ad,t}=\bar \rho_m \left( \frac{T_\mathrm{ad,t}}{T_m}
\right)^{1/(\gamma-1)}
~~~,~~~
\rho_\mathrm{ad,b}=\bar \rho_m \left( \frac{T_\mathrm{ad,b}}{T_m}
\right)^{1/(\gamma-1)}.
\label{eq:rho_ad}
\end{equation}
Consider now a parcel of gas near the top boundary. It has on average the
density $\bar \rho_t$. The density difference with the adiabatic profile, 
$\bar \rho_t - \rho_\mathrm{ad,t}$, accelerates the parcel through the volume.
The free fall velocity is estimated from a balance between the advection and
buoyancy terms, which reads in the non-dimensional variables used here
$|\rho (\bm v \cdot \nabla) \bm v| \sim \Delta \rho g d (d/\kappa_o)^2$,
where $\Delta \rho$ is the density difference of the moving parcel with the
adiabatically stratified background. The pressure variation experienced by the
falling parcel compresses the parcel by the same factor as the surrounding gas
(assuming the parcel does not exchange heat with its surroundings), so that
$\Delta \rho/\rho$ remains constant during the entire journey through the 
adiabatically stratified layer. It follows that $\Delta \rho/\rho$ keeps its
initial value of $(\bar \rho_t - \rho_\mathrm{ad,t})/\bar \rho_t$, and that the
square of the non-dimensional free fall velocity of the parcel arriving at
the bottom (which is expressed in units of ($\kappa_o/d)^2$) is
$\frac{\bar \rho_t - \rho_\mathrm{ad,t}}{\bar \rho_t} gd
\left(\frac{d}{\kappa_o}\right)^2$. 
A similar expression is derived if we start
the argument from the bottom boundary, so that we obtain two velocities,
$v_\mathrm{ff,t}$ and $v_\mathrm{ff,b}$ according to the formula

\begin{eqnarray}
v_\mathrm{ff,t}& = & \left( \frac{\bar \rho_t - \rho_\mathrm{ad,t}}{\bar \rho_t}
\mathrm{Pr} ~ \mathrm{Ra}\frac{T_o}{\Delta T} \right)^{1/2} ,
\nonumber
\\
v_\mathrm{ff,b} & = & \left( \frac{ \rho_\mathrm{ad,b} - \bar \rho_b}{\bar \rho_b}
\mathrm{Pr} ~ \mathrm{Ra}\frac{T_o}{\Delta T} \right)^{1/2}.
\label{eq:vff}
\end{eqnarray}

Fig. \ref{fig:Ekin_vff} verifies that one obtains with Eq. (\ref{eq:vff}) a
velocity representative of the convective velocity. The figure shows 
$E_{\mathrm{kin}}$ as a function of the energy density computed from the bottom
free fall velocity, $\rho_\mathrm{ad,b} v_\mathrm{ff,b}^2$. When this velocity
is small, the Reynolds number of the flow is too small for friction to be
negligible and the free fall velocity is a poor estimate of the true velocity.
For large velocities, there is a unique relation between $E_{\mathrm{kin}}$ and
$\rho_\mathrm{ad,b} v_\mathrm{ff,b}^2$ independent of any other parameters.

It is now expected that one obtains the same graph if one uses the kinetic
energy density computed at the top boundary, or equivalently, that
$\rho_\mathrm{ad,t} v_\mathrm{ff,t}^2 = \rho_\mathrm{ad,b} v_\mathrm{ff,b}^2$.
Fig. \ref{fig:vff_t_b} demonstrates that this is the case to within $\pm 20 \%$ for 
sufficiently large velocities and over three decades in
$\rho_\mathrm{ad,b} v_\mathrm{ff,b}^2$. The equality
of the two energy densities is the first important connection between the top
and bottom boundaries.

\begin{figure}
\includegraphics[width=8cm]{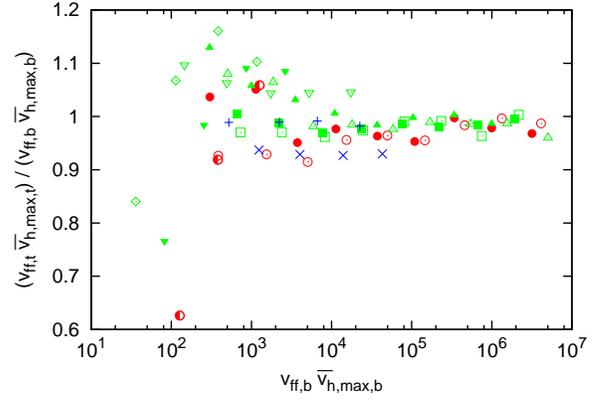}
\caption{(Color online)
$(v_\mathrm{ff,t}{\bar v}_{\mathrm{h,max,t}})/(v_\mathrm{ff,b}{\bar
v}_{\mathrm{h,max,b}})$ as a function of
$v_\mathrm{ff,b}{\bar v}_{\mathrm{h,max,b}}$
with the same symbols as in fig. \ref{fig:Nu_Ra}.}
\label{fig:vff_h}
\end{figure}

\begin{figure}
\includegraphics[width=8cm]{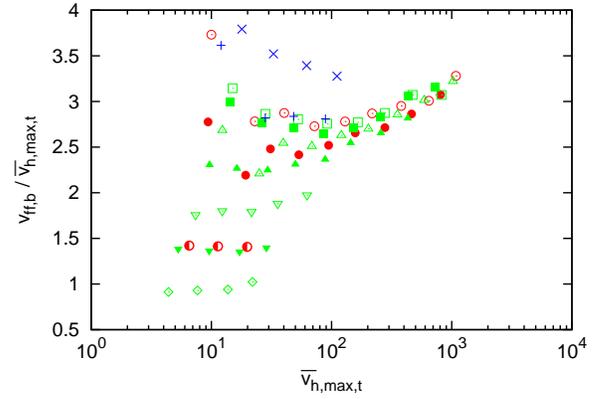}
\caption{(Color online)
$v_\mathrm{ff,b}/{\bar v}_{\mathrm{h,max,t}}$ as a function of ${\bar v}_{\mathrm{h,max,t}}$
with the same symbols as in fig. \ref{fig:Nu_Ra}.}
\label{fig:vff_vh}
\end{figure}

The second relation, shown in fig. \ref{fig:vff_h}, involves the local maxima of
horizontal velocity near the top and bottom boundaries visible in 
the bottom panel of fig. \ref{fig:profile}.
These maximum velocities, ${\bar v}_{\mathrm{h,max,t}}$ and 
${\bar v}_{\mathrm{h,max,b}}$, are listed in table \ref{table1}. 
According to fig. \ref{fig:vff_h}, they obey for sufficiently large velocities:
\begin{equation}
v_\mathrm{ff,t}{\bar v}_{\mathrm{h,max,t}} = v_\mathrm{ff,b}{\bar v}_{\mathrm{h,max,b}}.
\end{equation}

${\bar v}_{\mathrm{h,max,b}}$ is a free fall velocity computed from quantities evaluated
at the the bottom boundary, but it estimates the velocity of plumes arriving at
the top boundary. The typical velocity of the flow out of the arriving plumes is
the maximum average horizontal velocity, so that we expect
${\bar v}_{\mathrm{h,max,t}} \propto v_\mathrm{ff,b}$, and similarly
${\bar v}_{\mathrm{h,max,b}} \propto v_\mathrm{ff,t}$. The prefactors depend on the size
of the incoming plumes relative to the thickness of the layer in which the
outflow occurs. Fig. \ref{fig:vff_vh} shows that 
${\bar v}_{\mathrm{h,max,t}} \propto v_\mathrm{ff,b}$
approximately holds and that the prefactor indeed depends on
$d/H_o$ and $\Delta T/T_o$. The two proportionalities combine to
$v_\mathrm{ff,t}{\bar v}_{\mathrm{h,max,t}} \propto v_\mathrm{ff,b}{\bar v}_{\mathrm{h,max,b}}$.
It is remarkable that this combined relation is obeyed more accurately than the
two separate proportionalities, and that the proportionality factor in the
combined relation 
approaches 1 at high Rayleigh numbers as shown in fig. \ref{fig:vff_h}.

\section{Conclusion}
There are many ways how to depart from the Boussinesq approximation
\cite{Wu91,Kogan99,Burnis10,Ahlers07} and it is not known at present whether
there is anything universal about convection in general. 
The Prandtl number is a constant in a real gas and the Nusselt number depends
only on the Rayleigh number in the Boussinesq limit. Away from that limit,
several more control parameters determine the Nusselt number and it is a
challenge to find a suitable data reduction such that the $Nu(Ra)$ dependences
found for various density stratifications collapse on a single master curve,
which then must contain the Boussinesq limit. It has been shown in the present
paper that a good, if imperfect, data collapse is obtained if one introduces
into the scaling laws an effective density given by the geometric mean of the
maximum and minimum densities in the convecting layer. There is no theory
underpinning the relevance of the geometric mean. It seems likely that the
geometric mean will be of no help in some other cases of non-Boussinesq
convection, for example in liquids.

On the other hand, the importance of free fall velocities has been recognized
since long, and it has been shown here that two different estimates of free fall
velocities, based on quantities pertaining either to the top or the bottom boundary,
are related at high Rayleigh numbers
by the requirement that the kinetic energy densities computed from
the two velocities must be equal. The free fall velocities are also connected to
the total kinetic energy of the flow and the maxima of the horizontal velocity
profile. It will be worthwhile to check these relations in other forms of
convection.

\appendix*
\section{}
This appendix describes a few tests that have been performed in order to
validate the numerical code.
Direct validation of the code is problematic because published simulations of
compressible convection either focus on flow structures \cite{Toomre90,Porter00}
and are not useful as a benchmark, or are too close to the Boussinesq limit to
offer a stringent test \cite{Robins04}. A different route had therefore to be
taken.

First, the code simulating (\ref{eq:conti_BQ}-\ref{eq:T_BQ}) could be
validated against a completely independent spectral method \cite{Hartle03}. This
already verified most terms occurring in (\ref{eq:conti}-\ref{eq:T}). For
example, the spectral code calculates for a plane layer with no slip boundaries
in $z$, periodic boundary conditions in $x$ and $y$ with $l_x=l_y=2d$, and
$\mathrm{Pr}=0.7$ and $\mathrm{Ra}=1.6\times 10^4$ that $E_{\mathrm{kin}}=360$
and $\mathrm{Nu}=3.01$. If started from the initial conditions $\bm v=0$, 
$\theta= \sin (\pi z) \cos (2 \pi y) \cos (2 \pi x)$ at $t=0$, the system
(\ref{eq:conti_BQ}-\ref{eq:T_BQ}) needs to be integrated beyond $t=10$ to find
the final state. The new code with appropriately adapted boundary
conditions yields $E_{\mathrm{kin}}=351$ and $\mathrm{Nu}=2.96$ for 32 points
along $z$ and $64^2$ grid points in the $x,y-$plane. At twice that resolution in
each coordinate, the result is $E_{\mathrm{kin}}=357$ and $\mathrm{Nu}=2.99$.
The parameter $c^2$ in Eq. (\ref{eq:NS_BQ}) was set to $10^5$ so that the Mach
number is below $8.5 \times 10^{-2}$.

In a next
step, the code implementing (\ref{eq:conti}-\ref{eq:T}) was used to simulate the
propagation of sound waves. For this purpose, the term $- \hat{\bm
z}\mathrm{Pr}~ \mathrm{Ra}\frac{T_o}{\Delta T}$ was removed from
Eq. (\ref{eq:NS}). The remaining equations, if
linearized and neglecting dissipation, can be manipulated into
$$\partial_t^2 \bm v = \frac{T_o}{\Delta T} \frac{H_o}{d} \mathrm{Pr} ~
\mathrm{Ra}~
\nabla (\nabla \cdot \bm v).$$
This equation allows some simple analytical solutions. For example, there are
the eigenmodes depending only on $z$. The first of those eigenmodes for boundary
conditions imposing zero heat flux at the top and bottom boundaries is of the
form $v_z \propto \sin (\pi z) \cos (\omega t)$. This standing wave has a period of 
$\tau=2 \pi/\omega = 2 (\frac{T_o}{\Delta T} \frac{H_o}{d} \mathrm{Pr} ~
\mathrm{Ra})^{-1/2}$. For $\frac{\Delta T}{T_o}=0.1$, $\frac{d}{H_o}=0.01$,
$\mathrm{Pr}=0.7$ and $\mathrm{Ra}=2\times 10^4$, this predicts
$\tau=5.345 \times 10^{-4}$. With a resolution of 64 grid points along $z$, the
numerical result is $\tau=5.343 \times 10^{-4}$. One can similarly simulate
sound waves propagating in different directions. This is a
test for all terms involving $\nabla \cdot \bm v$. The
fact that simulations of convection at high $\mathrm{Ra}$ yield a temperature
gradient in the bulk close to the adiabatic gradient may also be regarded as a
test of the terms involving $\nabla \cdot \bm v$.

The dissipation rate of the standing wave of the previous paragraph can be used
to test the dissipative term. A more complete test is provided by the
energy budget which also tests the viscous heating term in
the temperature equation. If one takes the scalar product of Eq. (\ref{eq:NS})
with $\bm v$ and integrates Eqs. (\ref{eq:NS}) and (\ref{eq:T}),
multiplied by $\rho$, over all space,
one deduces from Eqs. (\ref{eq:conti}-\ref{eq:T}) that the time derivative of
kinetic plus internal energy is given by
$$\frac{d}{dt} \int \left[ \frac{1}{2} \rho \bm v^2 +
\mathrm{Ra}~\mathrm{Pr}~\frac{H_o}{d}\frac{1}{\gamma}
\frac{\rho}{\gamma-1}(T+\frac{T_o}{\Delta T}) \right] dV
=
G + V_1 +V_2$$
with

\begin{eqnarray}
G & = & -\mathrm{Pr} ~ \mathrm{Ra}\frac{T_o}{\Delta T} \int \rho v_z dV ,
\nonumber
\\
V_1 & = & \mathrm{Pr}  
\int v_i [\nabla^2 v_i + \frac{1}{3} \partial_i(\nabla \cdot \bm v)] dV ,
\nonumber
\\
V_2 & = & \mathrm{Pr} 
\int 2 [e_{ij}-\frac{1}{3} \nabla \cdot \bm v \delta_{ij}]^2 dV.
\nonumber
\end{eqnarray}

$G$ is the work done by gravity, $V_1$ the dissipated kinetic
energy and $V_2$ the heat generated through viscous dissipation. If we denote
time averages by angular brackets, we must find in a statistically stationary
state that $\langle G \rangle =0$ and $\langle V_1 \rangle + \langle V_2 \rangle
=0$. Since $V_1$ and $V_2$ have different forms and must be programmed
differently, the energy budget provides a good test for their correctness.
For the case $\frac{\Delta T}{T_o}=\frac{d}{H_o}=1$ and
$\mathrm{Ra}=6 \times 10^4$ included in table \ref{table1}, the simulations
yield $| \langle G \rangle / \langle V_1 \rangle| = 3.3 \times 10^{-3}$ and
$| \langle V_1 + V_2 \rangle / \langle V_1 \rangle| = 7.6 \%$ at a
resolution of $64^3$ grid points, and 
$| \langle G \rangle / \langle V_1 \rangle| = 4.2 \times 10^{-4}$ and
$| \langle V_1 + V_2 \rangle / \langle V_1 \rangle| = 4 \%$ at a
resolution of $128^3$ grid points. The formulae for $V_1$ and $V_2$ taken
together contain integrals of 18 derivatives, 6 of which are squared,
so that the typical
error on each derivative is about $0.1\%$. The volume integrals have been
computed by adding the integrands at each grid point, multiplied by the volume
of the cell surrounding each grid point. This method of integration is of first
order for general integrands \cite{Press86}, which explains why the error in
$| \langle V_1 + V_2 \rangle / \langle V_1 \rangle|$ is only halved when
doubling the resolution.

%\bibliography{convection}

\newpage
\clearpage

\begin{table}\centering
\begin{tabular}{ c c c c c c c c c c c c }

$\frac{\Delta T}{T_o}$ & $\frac{d}{H_o}$ & $\mathrm{Ra}$ & $\mathrm{Nu}$ &
$E_{\mathrm{kin}}$ & $\mathrm{Pe}$ & $100~ \bar \theta_m$ &
$\bar \rho_m$ & $\rho_\mathrm{max}$ & $\rho_\mathrm{min}$ &
${\bar v}_{\mathrm{h,max,b}}$ & ${\bar v}_{\mathrm{h,max,t}}$ \\

% include table.asc here

\hline 
0.1 & 0.12 & 2$\times 10^4$ & 1.19 & 48.3 & 9.59 & 0.09 & 1.05 & 1.10 & 1.00 & 10.3 & 10 \\ 
0.1 & 0.12 & 6$\times 10^4$ & 1.37 & 256 & 22.1 & 0.14 & 1.05 & 1.10 & 1.00 & 24.2 & 23 \\ 
0.1 & 0.12 & 2$\times 10^5$ & 1.55 & 941 & 42.3 & 0.19 & 1.05 & 1.10 & 1.00 & 43.5 & 40.4 \\ 
0.1 & 0.12 & 6$\times 10^5$ & 1.81 & 2.9$\times 10^3$ & 74.3 & 0.23 & 1.05 & 1.10 & 1.00 & 77.9 & 72.1 \\ 
0.1 & 0.12 & 2$\times 10^6$ & 2.18 & 9.32$\times 10^3$ & 133 & 0.21 & 1.05 & 1.10 & 1.00 & 138 & 129 \\ 
0.1 & 0.12 & 6$\times 10^6$ & 2.61 & 2.62$\times 10^4$ & 223 & 0.21 & 1.05 & 1.10 & 1.00 & 234 & 218 \\ 
0.1 & 0.12 & 2$\times 10^7$ & 3.32 & 8.24$\times 10^4$ & 396 & 0.23 & 1.05 & 1.10 & 1.00 & 410 & 379 \\ 
0.1 & 0.12 & 6$\times 10^7$ & 4.20 & 2.26$\times 10^5$ & 656 & 0.25 & 1.05 & 1.10 & 1.00 & 688 & 645 \\ 
0.1 & 0.12 & 2$\times 10^8$ & 5.49 & 6.52$\times 10^5$ & 1.11$\times 10^3$ & 0.17 & 1.05 & 1.11 & 1.00 & 1.16$\times 10^3$ & 1.08$\times 10^3$ \\ 
\hline 
1 & 1.2 & 2$\times 10^4$ & 1.22 & 50.8 & 8.2 & 1.67 & 1.49 & 2.01 & 1.02 & 11.6 & 9.38 \\ 
1 & 1.2 & 6$\times 10^4$ & 1.39 & 239 & 17.7 & 1.96 & 1.48 & 2.02 & 1.04 & 26.9 & 19.3 \\ 
1 & 1.2 & 2$\times 10^5$ & 1.59 & 808 & 32.4 & 2.09 & 1.49 & 2.02 & 1.04 & 48.9 & 31 \\ 
1 & 1.2 & 6$\times 10^5$ & 1.84 & 2.46$\times 10^3$ & 56.6 & 2.19 & 1.48 & 2.02 & 1.05 & 87.7 & 53.4 \\ 
1 & 1.2 & 2$\times 10^6$ & 2.24 & 7.9$\times 10^3$ & 102 & 2.05 & 1.49 & 2.03 & 1.04 & 157 & 94.6 \\ 
1 & 1.2 & 6$\times 10^6$ & 2.74 & 2.25$\times 10^4$ & 172 & 1.88 & 1.49 & 2.04 & 1.03 & 261 & 158 \\ 
1 & 1.2 & 2$\times 10^7$ & 3.46 & 7.08$\times 10^4$ & 305 & 1.93 & 1.49 & 2.05 & 1.02 & 452 & 278 \\ 
1 & 1.2 & 6$\times 10^7$ & 4.44 & 1.94$\times 10^5$ & 507 & 1.74 & 1.49 & 2.06 & 1.00 & 753 & 463 \\ 
1 & 1.2 & 2$\times 10^8$ & 5.96 & 5.62$\times 10^5$ & 863 & 1.46 & 1.49 & 2.07 & 0.99 & 1.28$\times 10^3$ & 805 \\ 
\hline 
10 & 12 & 2$\times 10^4$ & 1.28 & 40.6 & 3.49 & 4.52 & 5.71 & 11.05 & 1.52 & 13.7 & 6.55 \\ 
10 & 12 & 6$\times 10^4$ & 1.46 & 143 & 6.71 & 4.08 & 5.69 & 11.07 & 1.72 & 23.9 & 11.3 \\ 
10 & 12 & 2$\times 10^5$ & 1.69 & 511 & 12.8 & 4.22 & 5.67 & 11.09 & 1.96 & 45.2 & 19.9 \\ 
10 & 12 & 6$\times 10^5$ & 2.03 & 1.6$\times 10^3$ & 22.7 & 4.26 & 5.67 & 11.13 & 2.18 & 80.6 & \\ 
10 & 12 & 2$\times 10^6$ & 2.55 & 5.3$\times 10^3$ & 41.7 & 3.97 & 5.68 & 11.22 & 2.13 & 142 & \\ 
10 & 12 & 6$\times 10^6$ & 3.24 & 1.51$\times 10^4$ & 70.8 & 3.66 & 5.69 & 11.31 & 1.99 & 234 & \\ 
\hline 
\hline
0.1 & 0.1 & 2$\times 10^4$ & 1.45 & 108 & 14.5 & 0.11 & 1.03 & 1.07 & 1.00 & 15.5 & 15 \\ 
0.1 & 0.1 & 6$\times 10^4$ & 1.72 & 442 & 29.3 & 0.19 & 1.03 & 1.07 & 1.00 & 29.7 & 28.2 \\ 
0.1 & 0.1 & 2$\times 10^5$ & 2.09 & 1.56$\times 10^3$ & 54.9 & 0.22 & 1.03 & 1.07 & 1.00 & 55.8 & 52.6 \\ 
0.1 & 0.1 & 6$\times 10^5$ & 2.57 & 4.75$\times 10^3$ & 95.8 & 0.15 & 1.03 & 1.07 & 1.00 & 97.8 & 92.2 \\ 
0.1 & 0.1 & 2$\times 10^6$ & 3.27 & 1.5$\times 10^4$ & 170 & 0.43 & 1.03 & 1.07 & 0.99 & 177 & 165 \\ 
0.1 & 0.1 & 6$\times 10^6$ & 4.11 & 4.23$\times 10^4$ & 286 & 0.39 & 1.03 & 1.07 & 0.99 & 294 & 277 \\ 
0.1 & 0.1 & 2$\times 10^7$ & 5.41 & 1.26$\times 10^5$ & 494 & 0.38 & 1.03 & 1.08 & 0.99 & 517 & 472 \\ 
0.1 & 0.1 & 6$\times 10^7$ & 7.09 & 3.54$\times 10^5$ & 828 & 0.27 & 1.03 & 1.08 & 0.99 & 858 & 821 \\ 
\hline 
0.3 & 0.3 & 2$\times 10^4$ & 1.45 & 105 & 13.8 & 0.80 & 1.10 & 1.19 & 1.00 & 15.6 & 14.3 \\ 
0.3 & 0.3 & 6$\times 10^4$ & 1.72 & 422 & 27.7 & 1.00 & 1.10 & 1.20 & 1.01 & 30.3 & 26.3 \\ 
0.3 & 0.3 & 2$\times 10^5$ & 2.09 & 1.49$\times 10^3$ & 52 & 0.98 & 1.10 & 1.20 & 1.00 & 57.7 & 49 \\ 
0.3 & 0.3 & 6$\times 10^5$ & 2.58 & 4.57$\times 10^3$ & 91 & 1.07 & 1.10 & 1.21 & 0.99 & 103 & 85.9 \\ 
0.3 & 0.3 & 2$\times 10^6$ & 3.28 & 1.44$\times 10^4$ & 162 & 1.20 & 1.10 & 1.21 & 0.99 & 183 & 152 \\ 
0.3 & 0.3 & 6$\times 10^6$ & 4.16 & 4.06$\times 10^4$ & 271 & 1.12 & 1.10 & 1.22 & 0.98 & 305 & 253 \\ 
0.3 & 0.3 & 2$\times 10^7$ & 5.44 & 1.2$\times 10^5$ & 467 & 0.93 & 1.10 & 1.22 & 0.98 & 510 & 431 \\ 
0.3 & 0.3 & 6$\times 10^7$ & 7.10 & 3.24$\times 10^5$ & 768 & 0.89 & 1.10 & 1.22 & 0.97 & 844 & 724 \\ 
\hline 
1 & 1 & 2$\times 10^4$ & 1.44 & 88.4 & 11.6 & 2.38 & 1.29 & 1.61 & 1.04 & 15.2 & 12.3 \\ 
1 & 1 & 6$\times 10^4$ & 1.74 & 389 & 24.2 & 2.45 & 1.29 & 1.62 & 1.05 & 33.9 & 25 \\ 
1 & 1 & 2$\times 10^5$ & 2.10 & 1.24$\times 10^3$ & 43.2 & 2.68 & 1.29 & 1.63 & 1.03 & 58.8 & 39.6 \\ 
1 & 1 & 6$\times 10^5$ & 2.57 & 3.78$\times 10^3$ & 75.4 & 2.80 & 1.29 & 1.65 & 1.01 & 105 & 68.6 \\ 
1 & 1 & 2$\times 10^6$ & 3.26 & 1.18$\times 10^4$ & 133 & 2.74 & 1.29 & 1.66 & 1.00 & 185 & 120 \\ 
1 & 1 & 6$\times 10^6$ & 4.14 & 3.38$\times 10^4$ & 226 & 2.76 & 1.29 & 1.68 & 0.98 & 311 & 202 \\ 
1 & 1 & 2$\times 10^7$ & 5.48 & 1.03$\times 10^5$ & 396 & 2.52 & 1.29 & 1.69 & 0.96 & 537 & 354 \\ 
1 & 1 & 6$\times 10^7$ & 7.23 & 2.82$\times 10^5$ & 655 & 2.21 & 1.29 & 1.70 & 0.95 & 875 & 589 \\ 
1 & 1 & 2$\times 10^8$ & 9.75 & 8.09$\times 10^5$ & 1.11$\times 10^3$ & 1.81 & 1.30 & 1.71 & 0.94 & 1.52$\times 10^3$ & 1.03$\times 10^3$ \\ 
\hline

\end{tabular}
\caption{Summary of results. Listed are the control parameters
$\frac{\Delta T}{T_o}$,  $\frac{d}{H_o}$, and $\mathrm{Ra}$ together with
$\mathrm{Nu}$, $E_{\mathrm{kin}}$, $\mathrm{Pe}$, $\bar \theta_m$ multiplied by 100, $\bar
\rho_m$, $\rho_\mathrm{max}$, $\rho_\mathrm{min}$, ${\bar
v}_{\mathrm{h,max,b}}$, and ${\bar v}_{\mathrm{h,max,t}}$ (see text for
definitions). The table consists of three sections (corresponding to the color
code of the figures in the online version)
with different $\Delta T_{\mathrm{ad}}/\Delta T$, which is 4/5, 2/3 or 1/15
in going from the top to the end of the table. An entry is missing for
${\bar v}_{\mathrm{h,max,t}}$ if the profile of horizontal velocity has no maximum near
the top boundary.} 
\label{table1}
\end{table}

\newpage
\clearpage

\begin{table*}\centering
\begin{tabular}{ c c c c c c c c c c c c }

$\frac{\Delta T}{T_o}$ & $\frac{d}{H_o}$ & $\mathrm{Ra}$ & $\mathrm{Nu}$ &
$E_{\mathrm{kin}}$ & $\mathrm{Pe}$ & $100~ \bar \theta_m$ &
$\bar \rho_m$ & $\rho_\mathrm{max}$ & $\rho_\mathrm{min}$ &
${\bar v}_{\mathrm{h,max,b}}$ & ${\bar v}_{\mathrm{h,max,t}}$ \\

\hline 
3 & 3 & 2$\times 10^4$ & 1.42 & 61.6 & 8.11 & 4.08 & 1.75 & 2.56 & 1.17 & 13.5 & 9.66 \\ 
3 & 3 & 6$\times 10^4$ & 1.68 & 241 & 15.9 & 4.36 & 1.75 & 2.58 & 1.23 & 27 & 16.3 \\ 
3 & 3 & 2$\times 10^5$ & 2.05 & 853 & 30 & 4.51 & 1.75 & 2.63 & 1.21 & 53.6 & 29.4 \\ 
3 & 3 & 6$\times 10^5$ & 2.52 & 2.59$\times 10^3$ & 52.5 & 4.31 & 1.76 & 2.66 & 1.16 & 94.9 & 49.9 \\ 
3 & 3 & 2$\times 10^6$ & 3.25 & 8.8$\times 10^3$ & 96.7 & 4.29 & 1.76 & 2.70 & 1.11 & 177 & 88.4 \\ 
3 & 3 & 6$\times 10^6$ & 4.16 & 2.44$\times 10^4$ & 162 & 4.15 & 1.77 & 2.73 & 1.07 & 283 & 144 \\ 
3 & 3 & 2$\times 10^7$ & 5.59 & 7.57$\times 10^4$ & 286 & 3.83 & 1.77 & 2.76 & 1.04 & 496 & 257 \\ 
3 & 3 & 6$\times 10^7$ & 7.45 & 2.06$\times 10^5$ & 473 & 3.32 & 1.77 & 2.79 & 1.00 & 822 & 431 \\ 
\hline 
10 & 10 & 2$\times 10^4$ & 1.42 & 39.1 & 4.81 & 5.33 & 3.04 & 5.00 & 1.60 & 11.3 & 7.37 \\ 
10 & 10 & 6$\times 10^4$ & 1.70 & 149 & 9.39 & 5.09 & 3.04 & 5.05 & 1.84 & 22.1 & 12.3 \\ 
10 & 10 & 2$\times 10^5$ & 2.06 & 529 & 17.7 & 5.58 & 3.02 & 5.14 & 1.83 & 45.2 & 21.6 \\ 
10 & 10 & 6$\times 10^5$ & 2.58 & 1.65$\times 10^3$ & 31.3 & 5.41 & 3.03 & 5.22 & 1.73 & 78.3 & 35.6 \\ 
10 & 10 & 2$\times 10^6$ & 3.38 & 5.54$\times 10^3$ & 57.7 & 5.09 & 3.04 & 5.31 & 1.61 & 141 & 62.6 \\ 
10 & 10 & 6$\times 10^6$ & 4.40 & 1.61$\times 10^4$ & 98.8 & 4.82 & 3.05 & 5.38 & 1.52 & 237 & \\ 
\hline 
30 & 30 & 2$\times 10^4$ & 1.46 & 24.5 & 2.64 & 7.27 & 5.78 & 9.91 & 2.56 & 11.2 & 5.29 \\ 
30 & 30 & 6$\times 10^4$ & 1.77 & 97.6 & 5.46 & 5.66 & 5.74 & 10.03 & 3.32 & 19.5 & 9.58 \\ 
30 & 30 & 2$\times 10^5$ & 2.17 & 368 & 10.7 & 5.57 & 5.69 & 10.20 & 3.20 & 37.1 & 17.1 \\ 
30 & 30 & 6$\times 10^5$ & 2.73 & 1.14$\times 10^3$ & 18.9 & 5.37 & 5.70 & 10.36 & 2.99 & 65.8 & 28.7 \\ 
30 & 30 & 2$\times 10^6$ & 3.60 & 3.79$\times 10^3$ & 34.6 & 5.22 & 5.72 & 10.53 & 2.76 & 118 & \\ 
30 & 30 & 6$\times 10^6$ & 4.80 & 1.07$\times 10^4$ & 58.8 & 4.68 & 5.75 & 10.70 & 2.55 & 193 & \\ 
\hline 
100 & 100 & 2$\times 10^4$ & 1.50 & 19 & 1.6 & 6.86 & 12.29 & 21.72 & 5.17 & 8.92 & 4.4 \\ 
100 & 100 & 6$\times 10^4$ & 1.84 & 66.3 & 3.08 & 5.85 & 12.26 & 22.02 & 6.96 & 15.9 & 7.66 \\ 
100 & 100 & 2$\times 10^5$ & 2.33 & 242 & 5.95 & 5.56 & 12.21 & 22.38 & 6.57 & 29.8 & 13.7 \\ 
100 & 100 & 6$\times 10^5$ & 3.02 & 746 & 10.5 & 5.36 & 12.24 & 22.73 & 6.04 & 52.1 & 22 \\ 
100 & 100 & 2$\times 10^6$ & 4.03 & 2.47$\times 10^3$ & 19.2 & 4.92 & 12.28 & 23.14 & 5.53 & 92.5 & \\ 
100 & 100 & 6$\times 10^6$ & 5.34 & 7.25$\times 10^3$ & 33.1 & 4.41 & 12.29 & 23.53 & 5.11 & 161 & \\ 
\hline 
\hline
0.1 & 0.01 & 6$\times 10^3$ & 2.00 & 67.5 & 11.9 & -0.90 & 0.96 & 0.96 & 0.96 & 12 & 12 \\ 
0.1 & 0.01 & 2$\times 10^4$ & 2.90 & 392 & 28.6 & -0.44 & 0.96 & 0.96 & 0.96 & 28.2 & 28.1 \\ 
0.1 & 0.01 & 6$\times 10^4$ & 3.75 & 1.23$\times 10^3$ & 50.6 & -0.73 & 0.96 & 0.96 & 0.96 & 48.3 & 48.4 \\ 
0.1 & 0.01 & 2$\times 10^5$ & 5.10 & 4.14$\times 10^3$ & 92.7 & -0.40 & 0.96 & 0.96 & 0.96 & 90.3 & 88.9 \\ 
\hline 
1 & 0.1 & 2$\times 10^4$ & 2.26 & 155 & 20.6 & -1.67 & 0.73 & 0.74 & 0.72 & 18.3 & 17.9 \\ 
1 & 0.1 & 6$\times 10^4$ & 3.03 & 632 & 41.6 & -1.32 & 0.73 & 0.74 & 0.72 & 34.8 & 32.9 \\ 
1 & 0.1 & 2$\times 10^5$ & 4.11 & 2.24$\times 10^3$ & 78.2 & -1.47 & 0.73 & 0.74 & 0.72 & 66.3 & 62.1 \\ 
1 & 0.1 & 6$\times 10^5$ & 5.38 & 6.65$\times 10^3$ & 135 & -1.26 & 0.73 & 0.74 & 0.72 & 119 & 111 \\ 
\hline

\end{tabular}
\caption{Table \ref{table1} continued.}
\label{table2}
\end{table*}

\end{document}